\documentclass[reprint,superscriptaddress,amsmath,amssym,aps,prb]{revtex4-2}
\usepackage{bm}
\usepackage{graphicx}
\usepackage[usenames,dvipsnames]{color}
\usepackage[normalem]{ulem}
\usepackage[svgnames]{xcolor}
\usepackage{bm}
\usepackage{multirow}
\usepackage{float}
\usepackage{titlesec}
\usepackage[utf8]{inputenc}

\begin{document}

\title{Electronic thermal resistivity and quasi-particle collision cross-section in semi-metals}

\author{Adrien Gourgout}\thanks{Present address:Laboratoire des Solides Irradiés, CEA/DRF/lRAMIS, Ecole Polytechnique, CNRS, Institut Polytechnique de Paris, Palaiseau F-91128, France}
\affiliation{Laboratoire de Physique et d'Étude des Matériaux (ESPCI Paris - CNRS - Sorbonne Universit\'e) \\ PSL Research University, 75005 Paris, France}
\author{Arthur Marguerite}
\affiliation{Laboratoire de Physique et d'Étude des Matériaux (ESPCI Paris - CNRS - Sorbonne Universit\'e) \\ PSL Research University, 75005 Paris, France}
\author{Beno\^it Fauqu\'e}
\affiliation{JEIP, USR 3573 CNRS, Coll\`ege de France, PSL Research University, 11, Place Marcelin Berthelot, 75231 Paris Cedex 05, France}

\author{Kamran Behnia}
\affiliation{Laboratoire de Physique et d'Étude des Matériaux (ESPCI Paris - CNRS - Sorbonne Universit\'e) \\ PSL Research University, 75005 Paris, France}

\begin{abstract}

Electron-electron collisions lead to a T-square component in the electrical resistivity of Fermi liquids.  The case of liquid $^3$He illustrates that the \textit{thermal} resitivity of  a Fermi liquid has a T-square term, expressed in  m$\cdot$W$^{-1}$. Its natural units are $\hbar/k_FE_F^2$. Here, we present a high-resolution study of the thermal conductivity in bismuth, employing  magnetic field to extract the tiny electronic component of the total thermal conductivity and resolving signals as small as  $\approx 60 \mu$K. We find that the electronic thermal resistivity follows a T-square temperature dependence with a prefactor twice larger than the electric T-square prefactor. Adding this information to what has been known for other semi-metals, we find that the prefactor of the T-square thermal resistivity scales with the square of the inverse of the Fermi temperature, implying that the dimensionless fermion-fermion collision cross-section is roughly proportional to the Fermi wavelength, indicating that it is not simply set by the strength of the Coulomb interaction.  
\end{abstract}

\maketitle

\section{Introduction}

The collision rate between two  electrons, both confined to a thermal window at the Fermi level, increases with the square of temperature, leading to an electrical resistivity with quadratic temperature dependence. While this has been known for many decades \cite{Landau1936,Baber1937}, the microscopic mechanism by which these collisions degrade conduction has not been unambiguously identified.  The observation of T-square resistivity in dilute metals with a Fermi surface too small to allow Umklapp events \cite{Marel2011,lin2015,Wang2020} has initiated a regain of interest in this issue \cite{Kumar2021,Jaoui2021,Behnia2022,Jiang2023}.

A related topic is the amplitude of $A$ in this expression for the electric resitivity, $\rho$ of a Fermi liquid:
\begin{equation}
  \rho  = \rho_0 + A T^2
  \label{T-square1}
\end{equation}

Here, $\rho_0$, the residual resitivity is due to disorder and $A T^2$ is due to e-e scattering. The Kadowaki-Woods (KW) scaling \cite{Rice1968,Kadowaki1986,Tsujii2003,Hussey2005,Jacko2009}, which links $A$ (in $\mu\Omega$$\cdot$cm$\cdot$K$^{-2}$) to the Sommerfeld coefficient (the T-linear electronic specific heat dubbed $\gamma$) works in dense metals, where there is roughly one mobile electron per atom, but not in  dilute metals (where a mobile electron is shared by thousands of atoms) \cite{lin2015,Behnia2022}.

A T-square temperature dependence in the resistivity of semi-metallic bismuth was first observed in 1969 by Hartman \cite{Hartman1969}, who argued that it arises due to the Coulomb interaction  between carriers belonging to different valleys. This interpretation was subsequently contested by other authors \cite{Uher1977,kukkonen} who proposed an electron-phonon scattering mechanism tailored to produce a quadratic temperature dependence. A possible source of skepticism resided in the amplitude of the T-square resitivity prefactor in bismuth ($A=1.2 \times 10^{-8}\Omega$.cm.K$^{-2}$), which was many orders of magnitude larger than was observed in other metallic elements \cite{Rice1968,Uher2004,Behnia2022}. In Al, for example, it is more than 4 orders of magnitude smaller ($A=5.3\times 10^{-13}\Omega$.cm.K$^{-2}$) \cite{Garland1978}. Nevertheless, this large discrepancy is understandable nowadays \cite{Behnia2022}. Since the Fermi energy is much lower in bismuth than in Al, the e-e collision phase space is much larger in the former than in the latter.

\begin{figure*}[ht]
\begin{center}
\includegraphics[width=17cm]{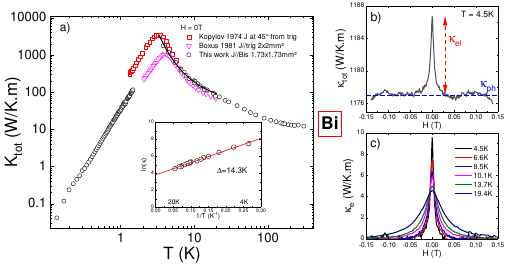}
\caption{a) Thermal conductivity of bismuth as a function of temperature in Bismuth. Red squares are taken from ref \cite{kopylov1973}, purple triangles are from ref \cite{Boxus1981} and black circles represent this work. The solid line is an exponential fit to the data in the $4 K < T < 20 K$ range. The inset shows the same data between 4K and plotted as ln($\kappa$) vs 1/$T$, demonstrating the exponential behavior in this temperature range with the solid red line representing a $\propto exp(\Delta/k_BT)$ fit to the data. b) Thermal conductivity as a function of magnetic field, applied along the trigonal axis, at $T$=4.5K. The plateau induced by the magnetic field represents the phononic contribution.  The difference between the zero-field conductivity and this plateau is the electronic contribution. c) The total thermal conductivity with the plateau subtracted as a function of magnetic field, at different temperatures. With increasing temperature, the electronic thermal conductivity decreases in amplitude and its field dependence broadens.}
\label{Fig1}
\end{center}
\end{figure*}

In this context, scrutinizing electronic thermal conductivity, $\kappa_e$, is insightful. The Wiedemann-Franz (WF) law establishes a correlation between the amplitude of thermal and electrical conductivities  by  the Sommerfeld value ($L_0 = \frac{\pi^2}{3} \frac{k_B^2}{e^2}=2.44\times10^{-8} V^2K^{-2}$).  Let us define the electronic thermal resistivity times temperature as $WT = \frac{L_0 T}{\kappa_{e}}$ \cite{Wagner1971}. It also displays a quadratic temperature dependence:

\begin{equation}
  WT = (WT)_0 + B T^2
  \label{T-square2}
\end{equation}

The first term is the  counterpart of $\rho_0$. The strict validity of the WF law at zero temperature implies that $(WT)_0=\rho_0$. On the other hand, a departure from this law is expected at finite temperature leading to $B > A$ (thanks to L$_0$, $A$ and $B$ have the same units). Thermal transport studies have quantified $B$ in Ni \cite{White1967}, W \cite{Wagner1971,Trodahl_1973}, Al \cite{Garland1978},  UPt$_3$ \cite{Lussier1994},  CeRhIn$_5$ \cite{Paglione2005}, WP$_2$ \cite{Jaoui2018}), Sb \cite{Jaoui2021}, and most recently in dilute metallic SrTi$_{1-x}$Nb$_x$O$_3$\cite{Jiang2023}. These studies confirmed the validity of the WF law in the zero temperature ($(WT)_0=\rho_0$)  and  detected a departure from it at finite temperature  giving rise an inequality between $B$ and  $A$. The thermal prefactor has been found to be larger than the electric prefactor. In other words, $B>A$, with the B/A ratio varying between 2 and 7 across different solids. The existence of a bound to this ratio has been a subject of theoretical debate \cite{Li-Maslov}. 

Decades ago, MacDonald and  Geldart \cite{MacDonald_1980b} argued that thermal resistivity caused by e-e scattering is simpler to understand than electric resistivity,  because of  the presence of Umklapp events. Experimentally, in contrast, thermal resitivity is more difficult to measure than electric resistivity. 

A discussion of thermal transport in Fermi liquids brings us to consider the case of normal liquid $^3$He \cite{Abrikosov_1959,Nozieres,wolfle1979, Vollhardt1990,Dobbs,Behnia2024}. Its thermal conductivity is proportional to the inverse of temperature at very low temperatures \cite{Wheatley1975,greywall1984}. Now, since $\frac{T}{\kappa}\equiv WT/L_0$ $(\kappa T)^{-1}$ and $B$ are proportional. It has been recently noticed that the evolution of $(\kappa T)^{-1}$ with pressure in $^3$He follows the scaling seen for the T-square resistivity with $E_F$ in metals. This implies that understanding  the amplitude of T-square thermal resistivity in a Fermi liquid, set by its Landau parameters, does not require Umklapp events or multiple Fermi pockets \cite{Behnia2022}.

Elemental bismuth has played a major role in the history of metal physics \cite{edelman1976,issi1979}. Its thermal conductivity has been the subject of several studies  \cite{kopylov1973,Boxus1981,PRATT197874,behnia2007}. However, they had little to say about the temperature dependence of the electron thermal resitivity, $WT$. This is not surprising, given the dominance of the phononic component of the thermal conductivity, $\kappa_{ph}$ in this semi-metal, in which one mobile electron is shared by 10$^5$ atoms. Since  $\kappa_{ph}$ is several orders of magnitude larger than $\kappa_{e}$, one needs a remarkably high resolution to extract the accurate temperature dependence of $\kappa_{e}$ and $WT$.

In this paper, we present a study of thermal conductivity in bismuth. By employing  magnetic field as a tool for separating $\kappa_{ph}$ and  $\kappa_{e}$, we extract the latter and find that the electronic thermal resistivity in bismuth follows a T-square temperature dependence and  $B$ is three time larger than $A$. In the second part of the paper, we compare the amplitude of $B$ and $A$ in several semi-metals and find that they are both proportional to the inverse of the square of the Fermi temperature. Then, we will compare the data to what is known about thermal transport in other Fermi liquid In S. I., the  quantity $(\kappa T)^{-1}$ is expressed in m$\cdot$W$^{-1}$. In a Fermi liquid,  when $\kappa$ refers to the fermionic thermal conductivity, the  natural unit of $(\kappa T)^{-1}$ is $\frac{\hbar}{E_F}{k_F}$. $(\kappa T)^{-1}$ is equivalent to $B$, save for their units, which arises because the former has not been multiplied by L$_0$. 

We show that the fermion-fermion dimensionless cross-section, dubbed $\zeta$ \cite{Behnia2022}, is less than unity in these semi-metals. This contrasts with $^3$He and strongly-correlated metals in which $\zeta \gg 1$. We also show that $\zeta$ steadily increases with carrier density, a feature hiding behind the empirical success of the extended KW scaling across various families of Fermi liquids \cite{Wang2020}.  

\begin{figure}[ht]
\begin{center}
\includegraphics[width=8.5cm]{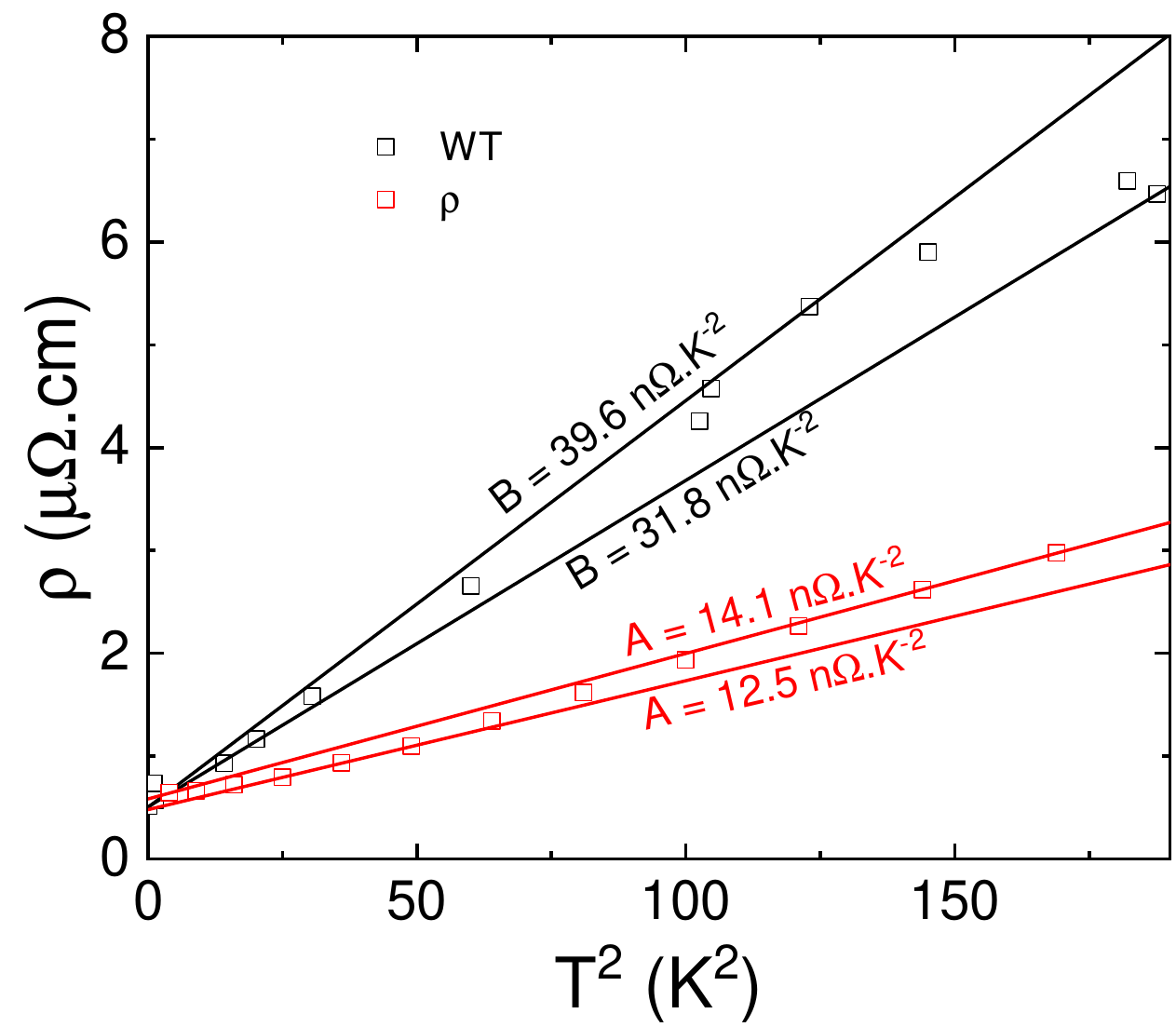}
\caption{Thermal (black squares) and electrical (red squares) resistivity as a function of the temperature squared. Thermal resistivity is expressed in units of $\mu\Omega$.cm using $L_0$ as a conversion factor. Solid lines are linear fits to the data with two extreme slopes. Within experimental margin, the intercepts are identical, but the slopes are different.}
\label{Fig2}
\end{center}
\end{figure}
\section{Results}
The bismuth single crystal used in this study was obtained commercially from Mateck. The crystal dimensions were $12 \times 1.73 \times 1.73 mm^3 $ with its longer dimension along the bisectrix crystalline axis. Its residual resistivity of 0.7 $\mu \Omega$.cm corresponds to a Room-temperature-Residual-resistivity ratio (RRR) of $\approx$230. This is lower than the RRR of larger crystals obtained from the same source and used in recent transport studies of bismuth \cite{Kang2022,Spathelf2022}. This size dependence of residual resistivity in bismuth indicates that at least a subset of carriers in mm-sized crystals are ballistic. Such a feature was also observed in other elemental semi-metals, such as antimony \cite{Jaoui2021}.
 
 Figure \ref{Fig1}a) shows the temperature dependence of the total thermal conductivity, $\kappa$$_{tot}$ in bismuth, at zero magnetic field, from room temperature down to low temperature ($T$=0.1K) and up to 300K. As seen in the figure, our data is close to what was previously reported by previous authors \cite{kopylov1973,Boxus1981}. We used a standard one-heater-two-thermometers method for measuring thermal conductivity in a dilution refrigerator and in a $^4$He cryostat (PPMS). We could not extract reliable data in the intermediate temperature range between 1.2 K and 3 K. This is the temperature range at which phonon thermal transport in bismuth displays hydrodynamic features, such as Poiseuille flow \cite{kopylov1973} and second sound \cite{Narayanamurti1972}. The breakdown of the Fourier heat flow picture in this region and the very large thermal conductivity of bismuth near this peak makes a reliable investigation difficult. The absence of data in this temperature range does not affect the results obtained and discussed in the present study. 
 
Above $\approx 20$ K, the temperature dependence is roughly $\kappa_{tot} \propto T^{-1}$. This is the so-called kinetic regime \cite{beck1974} where the lattice thermal conductivity is dominated by phonon-phonon Umklapp collisions \cite{berman1976thermal}. The scattering time is  longer than, but close to $\hbar/k_BT$, a generic feature of phonon heat transport near the Debye temperature \cite{Behnia_2019,Mousatov2020}.  Below 20 K, the increase in thermal conductivity with cooling becomes much faster than 1/$T$. As shown in the inset, in this temperature range $ln\kappa_{total}$ is a linear function of $T^{-1}$, indicating that thermal conductivity follows  $\kappa_{total} \propto exp(\frac{-\Delta}{T})$, with $\Delta=14.3$ K. This is the so-called Ziman regime \cite{beck1974}, where Umklapp collisions become rare. After peaking around 4 K, thermal conductivity starts to decrease and at low temperature, it recovers a T$^3$ temperature dependence corresponding to the Casimir \cite{beck1974,CASIMIR1938495} limit.

In order to separate the electronic and phononic contributions to the thermal conductivity, we focused on the variation of $\kappa_{total}$ a function of magnetic field following a procedure employed previously \cite{white1958,Uher1974,Jaoui2021,jaoui2022}. Thanks to the extremely large magnetoresistance of Bi \cite{Collaudin2015}, the electronic thermal conductivity is rapidly suppressed by a magnetic field which does not affect the phononic thermal conductivity. This can be seen in Figure \ref{Fig1}b, which shows the magnetic field dependence of the thermal conductivity at 4.5 K. Thermal conductivity is largest at zero magnetic field.  The application of a magnetic field of either polarity generates a plateau above  a small field of 0.025 T. This plateau represents the  phononic contribution to the total thermal conductivity, $\kappa_{ph}$. The difference between the amplitude of this plateau and the zero-field peak represents the electronic contribution  $\kappa$$_{el}$ at this temperature. Comparing the amplitude of the two components ($\kappa$$_{e}$ = 5 W/K.m and  $\kappa $$_{ph}$ = 1176 W/K.m at 4.5 K), gives an idea of the resolution required to separate the two components. We had to resolve a very small variation of an already small temperature difference. The temperature difference between the two thermometers measuring the temperature gradient was about 10 mK, which dropped by 60 $\mu$K, when then plateau was reached at 0.025 T.

Figure \ref{Fig1}c) shows the evolution of the field dependence of $\kappa_e=\kappa_{total}-\kappa_{ph}$ with temperature. The zero field peak gets broader as the temperature increases, which is consistent with what is expected given that warming, by reducing mobility, weakens orbital magnetoresistance. Thus, this procedure allows us to extract $\kappa$$_{e}$ as a function of temperature.

\begin{figure*}[ht]
\begin{center}
\includegraphics[width=17cm]{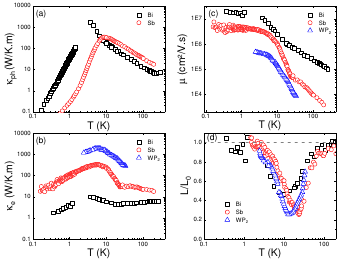}
\caption{a) Temperature dependence of the phononic thermal conductivity in Bi (black open squares) and Sb (red open squares). b) The electronic component of the thermal conductivity in Bi (black open squares), Sb \cite{Jaoui2021}(red open squares) and WP$_2$ \cite{Jaoui2018} (blue open squares).  c) The average mobility, $\mu$, extracted from $\kappa_e/T= L_0 (n+p) e \mu$ of the three semi-metals as a function of temperature. d) Normalized Lorenz number as a function of temperature in the three semi-metals. The WF law is verified at low and high temperature and there is a downward deviation at the intermediate temperature range.}
\label{Fig3}
\end{center}
\end{figure*}

Having quantified $\kappa_e$, we can now plot $WT= L_0 \frac{T}{\kappa_e}$ and compare it with electrical resistivity, $\rho$.  The result is shown in Fig. \ref{Fig2}, which presents the two resistivities, electrical and thermal, as a function of $T^2$. One can see that both are roughly linear in $T^2$ dependence with an identical intercept but different slopes. The prefactor of T-square electric resitivity ($A$ = 13 $\pm 1 n\Omega\cdot$cm$\cdot$K$^{-2}$) is slightly larger than what was reported by Hartman \cite{Hartman1969} ($A$ = 12 $n\Omega\cdot$cm$\cdot$K$^{-2}$) . On the hand, the prefactor of T-square electric resitivity ($B$= 35 $\pm$ 4 n$\Omega \cdot$ cm$\cdot$K$^{-2}$) had never been measured before. This is our main new experimental result. Thus, bismuth joins a club of materials in which a T-square thermal resistivity has been detected with a prefactor larger than the prefactor of electrical resistivity \cite{Behnia2022}. Fingers of both hands are sufficient for counting the members of this club.   

Note that the T$^2$  resistivity is restricted to temperatures below 20 K. Above this temperature, electrons are mostly scattered by phonons giving rise to different exponents for electrical and thermal resistivities \cite{Jaoui2018}. Note also that electron- phonon scattering becomes  elastic at high temperatures. This leads to a restoration of the Wiedemann-Franz law, as discussed below.

\section{Discussion}
It is instructive to compare bismuth with other semi-metals with a much larger carrier density. Figure \ref{Fig3} compares features of thermal transport in  Bi with Sb \cite{Jaoui2021,jaoui2022} and WP$_2$\cite{Jaoui2018}. In Figure \ref{Fig3}a), one sees a comparison of $\kappa$$_{ph}$ in Bi and Sb (In WP$_2$, thermal conductivity is dominated by $\kappa$$_{e}$ and  $\kappa$$_{ph}$ has not been extracted). In Bi and Sb, above 10 K, the amplitude and the behavior of $\kappa$$_{ph}$ is similar. The most remarkable difference between the two cases is the position of the peak, which occurs at a lower temperature in bismuth, which has a  lower Debye temperature.  
\begin{table*}[ht]
\centering
\begin{tabular}{|c|c|c|c|c|c|c|c|c|}
\hline
System & n=p (cm$^{-3}$)& A (n$\Omega$.cm/K$^2$) &$(\kappa T)^{-1} (m/W)$ &B (n$\Omega$.cm/K$^2$) & T$_F^{fermiology}$ (K) & T$_F^{n,\gamma}$ (K) &  $\gamma$ (mJ/mol/K$^2$)& References \\
\hline
Bi & 3.0$\times 10^{17}$ &12& 0.014 & 35 & 109 - 272 & 339 &  0.0085& \cite{Hartman1969,issi1979,Zhu2011}\\
\hline
Sb & 5.5$\times 10^{19}$&0.3& $2.9\times 10^{-4}$&0.6 & 844 - 931 & 1292 &  0.112& \cite{Jaoui2021,fauque2018,issi1979}\\
\hline
WTe$_2$&6.8$\times 10^{19}$ &  4.5 &$4.5\times 10^{-3}$& 11 & 250 - 500 & 484 &  6 &\cite{Zhu2015,xie2023puritydependent,CALLANAN1992627}\\
\hline
WP$_2$& 2.5$\times 10^{21}$& 0.017 &$3\times 10^{-5}$& 0.074 & 2130 - 6920 & 4109 &   2&\cite{Shonemann2017,Jaoui2018}\\
\hline
Mo & 1.4$\times 10^{22}$ &0.00148 &?& ? & 2070 - 16620 & 13821 &  1.9 &\cite{Desai1984,koelling1974}\\
\hline
W & 2.5$\times 10^{22}$ &8.7$\times 10^{-4}$ &$2.5\times 10^{-6}$& 6$\times 10^{-3}$ & 2950 - 21940 & 30288 &  0.84 &\cite{Wagner1971,Desai1984,GIRVAN19681485}\\
\hline
\end{tabular}
\caption{A comparison of semi-metals with an equal density of electrons and holes ($n=p$) spanning over four orders of magnitude of $n$. The table lists $n$, the electrical resistivity T$^2$ prefactor ($A$), the thermal resistivity prefactor ($(\kappa T)^{-1}$  and $B$, the latter is the former multiplied by L$_0$), the Fermi temperature (obtained by quantum oscillations or by the combination of $\gamma$ and $n$) and the Sommerfeld coefficient.}
\label{Table_1}
\end{table*} 

\begin{figure}[ht]
\begin{center}
\includegraphics[width=8.5cm]{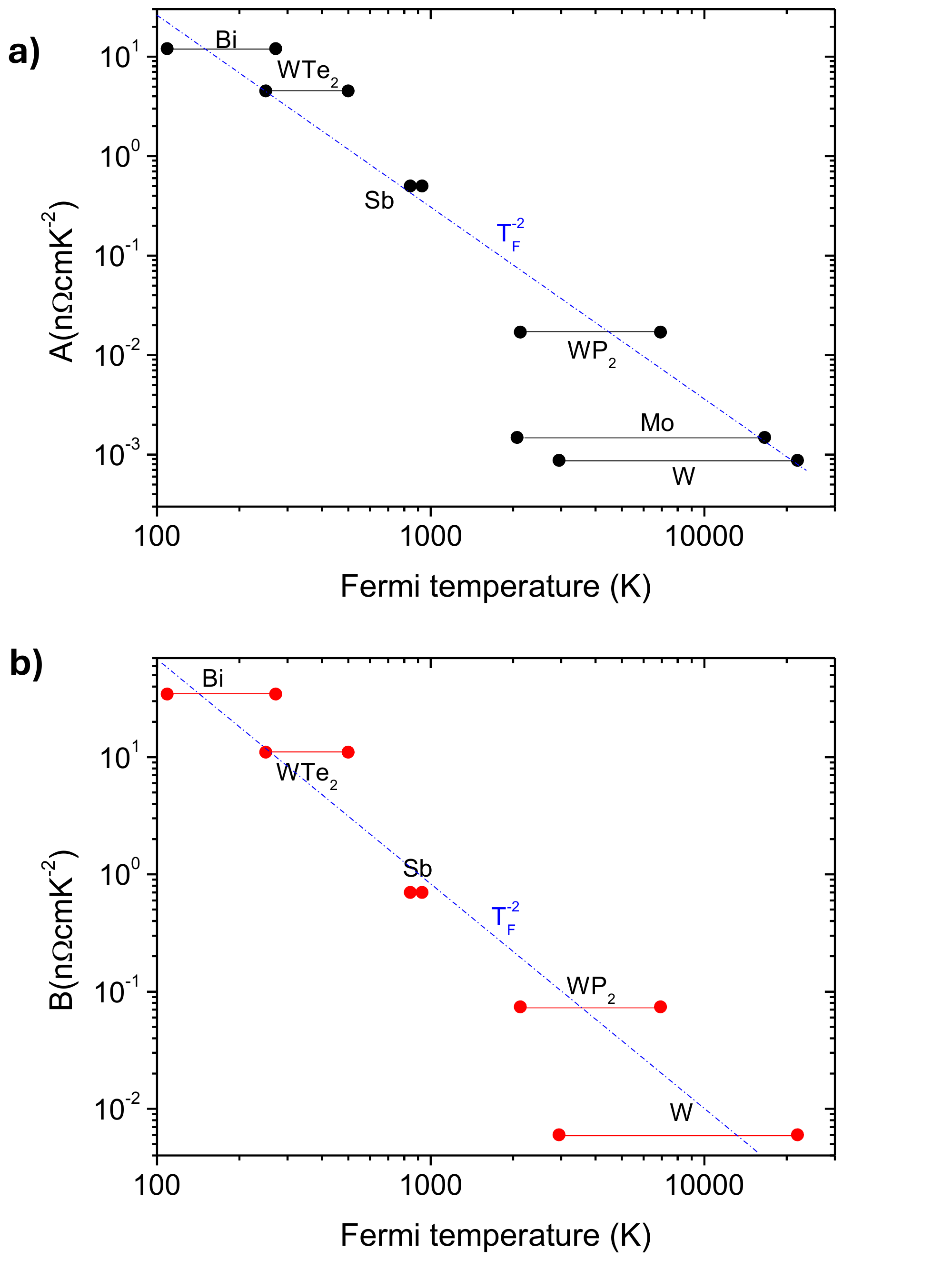}
\caption{a) The prefector of T-square electrical resitivity, $A$, as a function of the Fermi temperature in different semimetals. b) Same plot for the prefactor of T-square thermal resistivity, $B$ (in units of $n\Omega$.cm). In both plots, the dashed line represents a T$_F^{-2}$ slope. For each case,  the Fermi temperature has a minimum and a maximum value. They correspond to what was extracted in fermiology studies from the cross sections of the Fermi pockets and the effective masses found in the analysis of the period and the amplitude of quantum oscillations.}
\label{Fig4}
\end{center}
\end{figure}

Figure \ref{Fig3}b compares $\kappa_{e}$ in Bi, Sb and WP$_2$. It shows that the amplitude differs by three orders of magnitude, which is not surprising, given the large difference in their carrier concentration. As one can see in table \ref{Table_1}, the density of mobile electrons and holes is almost four orders of magnitude larger in WP$_2$ than in bismuth. In the case of Sb, $\kappa_{e}$ presents a sudden kink at $T \approx 20$ K. Below this temperature, scattering by phonons ceases to be a source of momentum loss by electrons \cite{jaoui2022}. A similar feature has been observed in WTe$_2$ \cite{xie2023puritydependent}, whose carrier density is comparable to antimony.   

Another relevant quantity extracted from the data is the electronic mobility. In the case of electrical conductivity, $\sigma$, mobility, $\mu$, is defined by the equation $\sigma=ne\mu$, where $n$ is the carrier density. Analogously, one can write for thermal transport: $\frac{\kappa_{el}}{T} = L_0 ne\mu_{th}$. Using the carrier density of each material (see table \ref{Table_1}). Figure \ref{Fig3}c compares the thermal mobility of carriers in the three semi-metals. One can see that mobility is largest in Bi, followed by Sb and finally WP$_2$, partially compensates with a higher electronic mobility. The large mobility of carriers in Bi, already known from electric resistivity measurements \cite{Hartman1969,Collaudin2015} is  a consequence of the fact that long wavelength electrons are less scattered by spatially confined disorder.

Figure \ref{Fig3}d  shows the temperature dependence of the normalized Lorenz  ratio ($L$/$L_0$) in the three semi-metals. The WF law is satisfied at both ends and there is a downward deviation at intermediate temperatures. The ratio becomes as low as 0.2 in Sb and in WP$_2$ and 0.4 in Bi. The curves look similar in the three cases, despite their absolute  (thermal and electrical) conductivities being orders of magnitude apart.

Note that the evolution  of the Lorenz ratio  with temperature is partially driven by electron-phonon scattering, which is inelastic at low temperature and contributes to keeping $L/L_0<1$. At high temperature, electron-phonon scattering becomes elastic and the normalized Lorenz  ratio tends towards unity.

The quantification of the electronic thermal transport in bismuth permits us to draw a global picture of the evolution of the T-square prefactors in four elemental semi-metals (Bi, Sb, Mo, W) together with two newly discovered Weyl semi-metals (WTe$_2$ and WP$_2$). Table \ref{Table_1} lists their T-square prefactors together with their other physical properties. We note that the non-trivial topology of electronic structure in WP$_2$ or in WTe$_2$ does not lead to any visible effect. This is presumably because  T-square resistivity arises by applying the Pauli Exclusion Principle twice (to both colliding electrons). The states which are out of the thermal window centered at the Fermi level do not play a role. Therefore, band crossings, which occur deep in the Fermi sea are not relevant to the Fermi liquid transport \cite{Haldane2004}.

Since these metals have several Fermi pockets with nontrivial morphologies, there is no unique Fermi temperature. The table lists the results of two procedures. In the first, minimal and maximal values of the Fermi temperature have been calculated by taking the effective mass and the frequency of quantum oscillations obtained  for each Fermi  pocket in fermiology studies. In the second, the average Fermi temperature has been extracted using the one-band expression which links, $\gamma$ to T$_F$ through the carrier density ($\gamma=2V_m\frac{\pi}{2}k_B\frac{n}{T_F}$), where $V_m$ is the molar volume and the factor 2 is due to the presence of a hole density as large as the electron one.

One can immediately notice that the ordinary KW scaling ($A \propto \gamma^2$) does not operate here \cite{Wang2020,Behnia2022}. To see this, it suffices to compare bismuth with tungsten. While $A$ is four orders of magnitude larger in Bi, $\gamma$ is a hundred times larger in W. This is not surprising. Given the wide variety of their carrier density, one cannot employ the common version of the KW scaling \cite{Behnia2022}. Indeed, the electronic specific heat depends both on carrier density and the degeneracy temperature. For the same reason the $A/\gamma$ scaling fails \cite{mccalla2019} in the case of dilute metallic strontium titanate. 

On the other hand, an `extended' version of the KW scaling, a proportionality between $A$ and  T$_F^{-2}$, has been shown to be relevant to a wide variety of Fermi liquids \cite{lin2015,Wang2020}. Figure \ref{Fig4} shows that this is indeed the case here. Panel a) is a log-log plot of $A$ in these semi-metals as a function of their Fermi temperature (see also Figure 3 in ref. \cite{Behnia2022} and Fig.12 in ref. \cite{Nakajima2024} ). Panel b shows the same for $B$. For both prefactors, the available data scatters around a line representing T$_F^{-2}$. The large extension of the Fermi temperatures indicates that while this correlation is robust, it should not be taken too seriously.

Since their conception decades ago, the slope seen in such plots has escaped a quantitative explanation. The scaling plots by Rice \cite{Rice1968} and by Kadowaki and Woods \cite{Kadowaki1986} between $A$ and $\gamma^2$ yielded a quantity expressed in $\Omega$.m.J$^{-2}$.K$^{4}$.mol$^2$ and not easy to decipher. Plotting $A$ a s function of Fermi temperature \cite{lin2015,Wang2020}  clarified the situation to some extent. Invoking the quantum of conductance, could write $A$ as: 

\begin{equation}
 A= \frac{\hbar}{e^2}\frac{\ell_{quad}}{T_F^2}
  \label{A-E_F}
\end{equation}

Taking this equation as a guide, the slope of $A$ \textit{vs.} $T_F^2$ yields a length scale, $\ell_{quad}$. In the case of elemental semi-metals, what was noticed first by Rice \cite{Rice1968} corresponds to $\ell_{quad}$=1.6nm. In the case of strongly correlated metals, what was popularized by Kadowaki and Woods \cite{Kadowaki1986} corresponds to $\ell_{quad}$=40 nm. Wang \textit{et al.} showed that available data for all Fermi liquids, suggest a link between $A$ and E$_F$ leading to a $\ell_{quad}$ between these two extreme values \cite{Wang2020}. Our data for both $A$ and for $B$ is compatible with this observation. Both plots lead to an $\ell_{quad}$ of the order of a few nanometers. It remains to figure out what hides behind the amplitude of $\ell_{quad}$.

\begin{figure}[ht]
\begin{center}
\includegraphics[width=8.5cm]{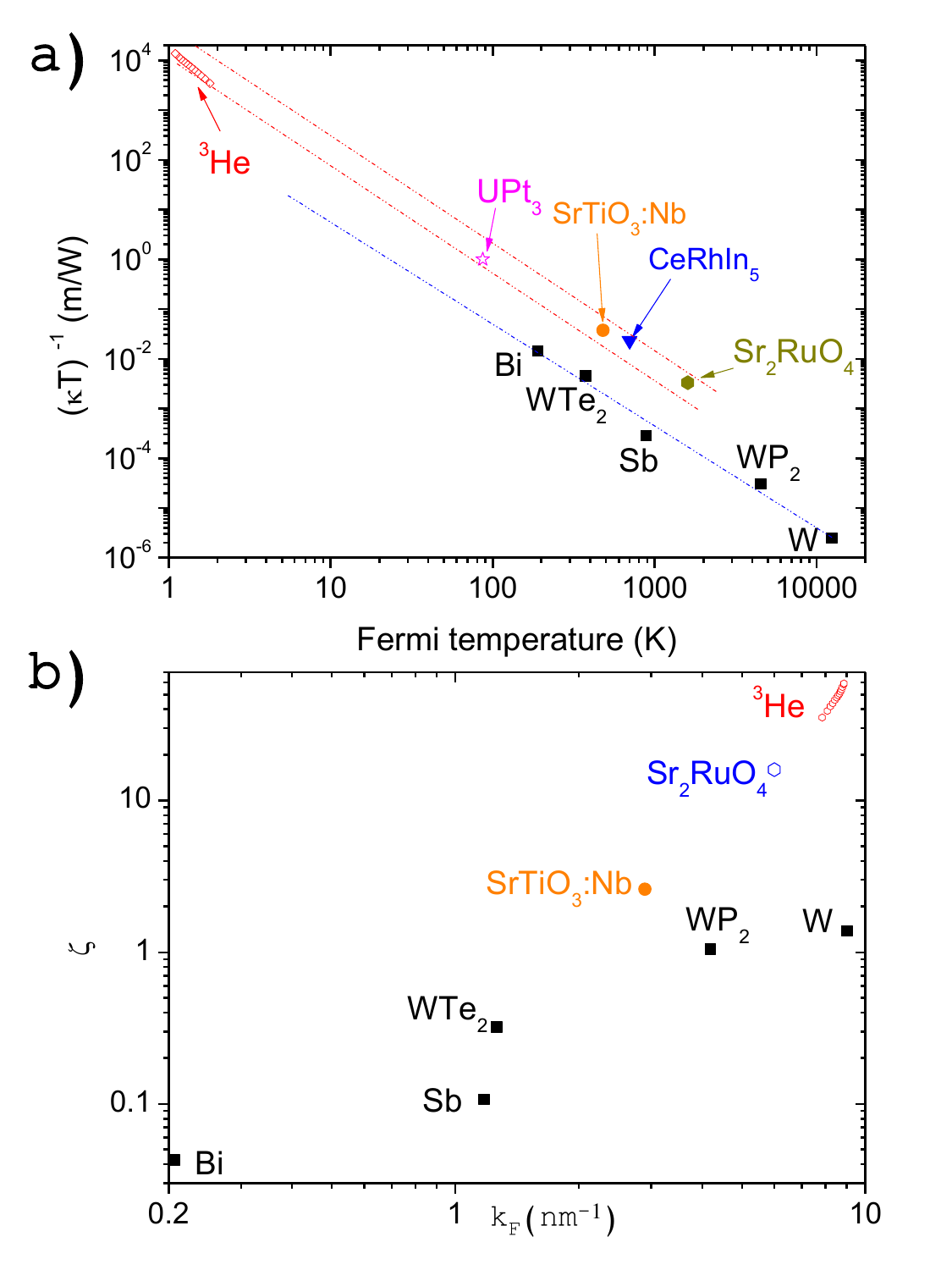}
\caption{a) The prefactor of T-square thermal resistivity in units of mW$^{-1}$ as a function of Fermi temperature in semimetals. Also included are three correlated metals(UPt$_3$ \cite{Lussier1994,Joynt2002}, CeRhIn$_5$ \cite{Paglione2005,Hegger2000}, and Sr$_2$RuO$_4$ \cite{Maeno1997,Mackenzie2003,Hassinger2017}), a dilute metal (Nb-doped strontium titanate \cite{Jiang2023,collignon2019}), as well as the normal liquid $^3$He under pressure \cite{greywall1984}. b) The dimensionless fermion-fermion cross section $\zeta$ in elemental semi-metals as a function of their wave-vector.$\zeta$ remains below unity and displays a tendency to enhance with increasing k$_F$. As shown in the case of  $^3$He and two other metals $\zeta$ can exceed unity. The largest known $\zeta\approx 60$ occurs for $^3$He at the onset of solidification.}
\label{Fig5}
\end{center}
\end{figure}

The case of normal liquid $^3$He sheds light on this issue.  Its thermal conductivity is proportional to the inverse of temperatures as a result of fermion-fermion scattering \cite{Wheatley1975,greywall1984} (but only at very very low temperatures \cite{Behnia2024}). The amplitude of this residual $\kappa T$ term has been carefully measured as a function of pressure up to the threshold of solidification \cite{greywall1984}. This leads to the quantification of the fermion-fermion scattering time, $\tau_{\kappa}$, which is linked to the Fermi energy by \cite{Behnia2022,Behnia2022b}:

\begin{equation}
\tau_{\kappa}= \frac{1}{\zeta} \frac{\hbar E_F}{(k_BT)^2}
  \label{zeta}
\end{equation}

Here, $\zeta$ is a dimensionless parameter, which quantifies the strength of collision. Often, it is assumed  to be close to unity (See for example, equation 13 of ref. \cite{Poduval2023}, where it is assumed that $\zeta= \frac{\pi}{4}$). Now, using the expression linking the thermal conductivity and the specific heat per volume ($\kappa=\frac{1}{3}$Cv$_F^2\tau_{\kappa}$), one finds:

\begin{equation}
(\kappa T)^{-1}= \frac{9}{2} \frac{\hbar}{k_FE_F^2}\zeta
  \label{kappaT}
\end{equation}

Thus, the natural units of $(\kappa T)^{-1}$ is $\frac{\hbar}{k_FE_F^2}$, a combination of the Fermi energy and the Fermi wave-vector. Its amplitude is not merely set by the inverse of the square of the Fermi energy $E_F^{-2}$, but also by the  $\zeta/k_F$. This ratio is sets the empirical length scale $\ell_{quad}$. 

With these considerations in mind, we have plotted in Figure \ref{Fig5}a the variation of $(\kappa T)^{-1}$ (in units of W/m) as a function of their average Fermi energy (in units of K), estimated from the carrier density and Sommerfeld coefficient of each solid. For comparison, the figure also contains the data for $^3$He, and a number of correlated metals as well as dilute metallic strontium titanate. Clearly, the correlation seen for the weakly correlated semimetals does not hold for other cases, indicating the non-universality of $\zeta$ and implying that the collision strength is not identical across various metallic families. 

Figure \ref{Fig5}b shows the extracted $\zeta$, extracted from thermal transport data, as a function of the average Fermi wavelength. The first feature to notice is that $\zeta$ in all semimetals is of the order of unity or less, as one may expect. What is more mysterious is the apparent correlation between $\zeta$ and $k_F$. We can now see that the success of the 'extended' Kadowaki-Woods scaling is due to a rough linearity between  $\zeta$ and $k_F$. At present, there is no satisfactory understanding of this feature. A larger Fermi radius may enhance the collision cross section in momentum space, but it will reduce it in the real space by making the Fermi wavelength smaller. 

As seen in the figure, in strongly correlated metals and in $^3$He, $\zeta$ is much larger than unity. The largest known $\zeta$ is attained by $^3$He abutting solidification \cite{Behnia2022}. $^3$He atoms are known to interact through a refined version of Lennard-Jones potential \cite{godfrin2022dynamics}. This is a short-range and, given the distance between atoms, and attractive interaction. In contrast, the screened Coulomb interaction between electrons is long-range and repulsive. It presumably dominates in strongly correlated metals. However, electrons can also interact attractively by exchanging phonons. Thanks to superconductivity, this is a widely known, although not quantitatively domesticated, feature of electrons in metals.

In this context, it is instructive to examine how  strontium titanate fits in this picture. The magnitude of the electric permittivity in this quantum para-electric solid is four orders of magnitude larger than vacuum permittivity \cite{Muller1979}. Therefore, the screening of the Coulomb interaction is very large. Nevertheless, as seen in Figure \ref{Fig5}b, its $\zeta$ is larger than what is found for semi-metals. This indicates that what drives electron-electron collision rate may not be Coulomb interaction alone. Exchange of phonons between electrons plays play a significant role in generating  T$^2$ thermal resistivity in Sb \cite{jaoui2022}. Theoretical considerations on the origin of electron hydrodynamics in WTe$_2$ \cite{Vool2021} identified phonon exchange as  a driver. Interestingly, early theoretical attempts in understanding the amplitude of T-square electric resitivity in Al \cite{Macdonald1980} and in noble metals \cite{Macdonald1981} indicated a crucial role played by phonon exchange. 

The threefold difference in $\zeta$ between Sb and WTe$_2$, despite their similar carrier density is noteworthy. Carriers are heavier and the Fermi velocity is lower in WTe$_2$ than in Sb. This van give rise to a higher dimensionless  potential-to-kinetic ratio in WTe$_2$ and play a role in its larger $\zeta$.

Quantifying the amplitude of e-e scattering and T$^2$  resistivity in real metals in which the electronic band structure, the pseudo-potential map and the phonon spectrum are all known is a challenge to computational condensed matter physics. We note a recent dynamical mean field theory study  devoted to SrVO$_3$ \cite{abramovitch2024respective}.

In summary, we measured thermal conductivity of elemental bismuth and quantified its electronic and the phononic components.  The Wiedemann-Franz is obeyed at zero temperature, but a downward deviation appears at finite temperature. By combining our result with what has been reported for other semi-metals, we drew a picture of the evolution of the amplitude of the prefactors of the quadratic resistivities. The dimensionless cross section of fermion-fermion scattering seems to grow linearly with the Fermi wave-vector and appears not set by Coulomb interaction. 

\section{Acknowledgements}
We thank Mikhail Feigel'man, Yo Machida  and Zengwei Zhu for stimulating discussions. This work was supported by a grant from the \^Ile de France Region. KB acknowledges a stay in KITP, where this paper was partly written, supported  by the National Science Foundation under Grant Nos. NSF PHY-1748958 and NSF PHY-2309135.

\bibliography{ref}
\end{document}